\documentstyle[aps,floats,epsf,twocolumn]{revtex}
\begin{document} \draft

\title{Quasiparticle spectra in the vicinity of a $d$-wave vortex.}

\author{M. Franz and Z. Te\v{s}anovi\'c
}
\address{Department of Physics and Astronomy, Johns Hopkins University,
Baltimore, MD 21218
\\ {\rm(\today)}
}
%
\address{~
\parbox{14cm}{\rm
\medskip
We discuss the evolution of the local quasiparticle spectral density 
and the related tunneling conductance  
measurable by the scanning tunneling microscope, as a function of distance $r$
and angle $\theta$ from the vortex core in a $d_{x^2-y^2}$ superconductor. 
We consider the effects of electronic disorder and of a 
strongly anisotropic tunneling matrix element, and show 
that in real materials they will likely obscure the 
$\sim 1/r$ asymptotic tail in the zero-bias tunneling conductance 
expected from the straightforward semiclassical analysis. 
We also give a prediction for the tunneling conductance 
anisotropy around the vortex core and 
establish a connection to the structure of the tunneling matrix 
element. 
}} 
\maketitle


%

\section{Introduction.}

While there remains almost no doubt at present that the hole-doped high-$T_c$
cuprate superconductors posses an unconventional $d_{x^2-y^2}$ order
parameter, the microscopic origin and many phenomenological consequences 
of this fact remain to be 
understood. The situation is most pressing in the presence of applied 
external magnetic field where the interplay between the spatially varying 
order parameter, supercurrents, and low energy quasiparticles results in a
great variety of novel effects. In the Meissner state Yip and Sauls 
had predicted a nonlinear Meissner effect\cite{ys} manifested by an anisotropic
component of the in-plane penetration depth linear in field, a unique
consequence of the nodal structure of the $d$-wave order parameter. However,
despite considerable experimental effort, this effect has not been clearly
observed in cuprates\cite{ysexp} and, very recently, Li, Hirschfeld, W\"{o}lfle
\cite{li1} argued that the effect might not be observable in cuprates, even in 
principle, due
to the highly non-local nature of the electromagnetic response of a
$d$-wave superconductor at low temperatures. In the mixed state
similar non-local effects have been predicted to result in a very rich
equilibrium vortex lattice structure phase diagram\cite{franz1}, but again,
no conclusive experimental confirmation has yet been reported. 

Among the more successful theoretical predictions specific to the 
$d$-wave order
parameter is Volovik's prediction of a $\sim T\sqrt{H}$ contribution to 
the specific heat\cite{volovik1} which was identified in measurements on 
YBa$_2$Cu$_3$O$_7$ (YBCO)
single crystals\cite{moler1}. Although subsequent experimental investigations
reported deviations from the precise $\sqrt{H}$ behavior\cite{revaz}, 
they were 
consistent with more general scaling relations\cite{simon1,volovik2} based
on the same general physical picture. In its simplest form the 
$\sim T\sqrt{H}$ behavior can be derived by observing that the supercurrent
orbiting each individual vortex Doppler-shifts the local quasiparticle 
spectrum, $E_{\bf k}\to E_{\bf k}-{\bf v}_s({\bf r})\cdot{\bf k}$, which 
in turn results in 
finite residual density of states
 $N(E=0,{\bf r})\propto v_s({\bf r})\propto 1/r$. 
Integrating $N(0,{\bf r})$ up to the inter-vortex distance $R_H\propto 
1/\sqrt{H}$
one obtains the total density of states per vortex and the Volovik's result
then follows on multiplying by the number of vortices $n_v=H/\Phi_0$, with 
$\Phi_0$ the flux quantum. We note that behavior consistent with the 
$\sqrt{H}$ dependence of the residual density of states was found in
numerical calculations within the  Bogoliubov-de Gennes (BdG) formalism
\cite{wang1}.

Here we wish to theoretically address the surprising 
fact that the expected $N(0,{\bf r})\propto 1/r$ 
dependence of the local density of states (LDOS)
 {\em has not} been observed in the scanning tunneling
spectroscopy (STS) 
measurements\cite{renner1,renner2,fischer1}, despite the fact
that this state of the art technique definitely possesses the required 
spatial and energy resolution. Inspection of the data reveals that 
instead of a $1/r$ asymptotic tail at large distances the spectra 
for YBCO and Bi$_2$Sr$_2$CaCu$_2$O$_8$ (BSCCO) recover their zero-field 
profiles within short distances from the cores on the order of several 
coherence lengths, beyond which the spectra remain unchanged. There are several
compelling reasons why is it desirable to resolve this potential conflict 
between the STS and thermodynamic measurements. The most important one has to 
do with confirming  the picture of a well defined $d$-wave quasiparticle in 
the superconducting state of cuprates. In particular, since it is  
generally believed that the normal state of the cuprates is {\em not} a 
conventional Fermi liquid, it is of considerable importance to verify, 
in as exhaustive detail as possible, that the ordered state below $T_c$, where
the Fermi liquid scenario is believed to apply, indeed exhibits all the
expected features predicted by the theory. When this fundamental issue
has been clarified one can perhaps hope to tackle greater problems in the 
field, such
as the nature of the non-Fermi liquid like behavior above $T_c$ and the 
origin of the pairing mechanism. 

In order to achieve our goal we first demonstrate, by direct comparison
to the results of fully self-consistent BdG theory for a
single vortex, that the semiclassical Volovik approach indeed captures the 
right physics of single-particle excitations {\em outside} the vortex core. 
We then proceed to extend this approach to the realistic case of quasiparticles
with finite lifetimes and with strongly anisotropic $c$-axis tunneling
matrix element. Our main result is that the above mentioned 
STS data away from the vortex core can be
understood by considering the effect of the matrix
element $M_{\bf k}$ for the electron tunneling along the $c$ axis. 
According to the band structure considerations for tetragonal cuprates 
$M_{\bf k}$
exhibits the same anisotropy as the gap function, i.e. it vanishes 
linearly on the Fermi surface near the zone diagonals
\cite{wheatley1,andersen1} . In such a case the 
tunneling conductance $g(V,{\bf r})$ is {\em not} simply proportional 
to the temperature-broadened LDOS, but reflects the additional structure in 
$M_{\bf k}$. Analysis of a case when $M_{\bf k}\propto\cos 2\phi$
leads to a surprising conclusion that the 
the power law in the decay of the zero-bias  
tunneling conductance $g(0,{\bf r})$ changes to $1/r^3$, meaning that it
vanishes much faster  than the expected $1/r$ decay for LDOS.
We argue that combined with the effects of electronic disorder, 
which tends to further wash out the effect of supercurrents on $g(0,{\bf r})$,
 this mechanism is responsible for the absence of the $1/r$ conductance tail 
observed in STS, {\em despite} 
the fact that LDOS itself exhibits the $1/r$  behavior.

We further show that there exists a direct relationship 
between the structure of $M_{\bf k}$ and the real space anisotropy of 
$g(V,{\bf r})$ around the vortex core. In particular we predict that 
the tunneling conductance along the circle centered at the core
will exhibit fourfold anisotropy with maxima along the nodes and minima
along the antinodes.  In contrast, the angular dependence of LDOS is 
exactly opposite, with minima along the nodes and maxima
along the antinodes. This prediction is directly testable by the STS
experiment and the result should shed light on the structure of the tunneling 
matrix element $M_{\bf k}$ which is of considerable interest in its own right.
In the absence of disorder the ratio of minimum to 
maximum is $\sqrt{2}$ for both $g(0,{\bf r})$ and LDOS; disorder will
reduce this value by adding an approximately equal amount of constant
conductance to both the maximum and the minimum.

\section{Tunneling conductance: general formalism}

\subsection{Zero field}

Tunneling conductance at bias $V$ between a superconductor and a 
normal metal is given by

\begin{equation}
g(V)=-\int_{-\infty}^{\infty}d\omega f'(\omega-eV)\sum_{\bf k} 
|M_{\bf k}|^2 A({\bf k},\omega),
\label{cond1}
\end{equation}
where $f$ is the Fermi function and $A({\bf k},\omega) =-2{\rm Im}
\cal G({\bf k},\omega)$ is the spectral 
function of a superconductor related to the diagonal part of the full
superconducting Greens function $\hat G({\bf k},\omega)$. 
In quasi-2D cuprates the spectral function is taken to describe 
electrons within a single $ab$-plane and, correspondingly, ${\bf k}$ refers 
to a 2D wave-vector. While Eq.\ (\ref{cond1}) could be written down on purely 
intuitive grounds a more detailed discussion of how one actually implements
this dimensional reduction is given in the appendix.

For the clean system the formulation based
on Eq.\ (\ref{cond1}) is fully equivalent to the analogous expression in terms 
BdG wave-functions but allows for a straightforward inclusion
of the effects of finite quasiparticle lifetime. In the absence of field the
diagonal Greens function assumes the well known form\cite{mahan1} (taking
$\hbar=1$) 
\begin{equation}
{\cal G}({\bf k},\omega)={(\omega-i\Gamma)+\epsilon_{\bf k}\over
(\omega-i\Gamma)^2-\epsilon_{\bf k}^2-\Delta_{\bf k}^2},
\label{g1}
\end{equation}
where $\epsilon_{\bf k}$ is the normal state electron dispersion, 
$\Delta_{\bf k}=\Delta_d\cos2\phi$ is the $d$-wave gap, and $\Gamma$ models 
the quasiparticle lifetime broadening that 
results from random disorder and inelastic
processes. In a $d$-wave superconductor, strictly speaking, the lifetime 
effects should be described by a frequency and wave-vector dependent self 
energy $\Sigma({\bf k},\omega)$ whose precise structure, however, 
is not well understood 
at present. Since we are mainly interested in the effect of 
vortices on the spectral properties at the lowest energies, we shall in the 
following ignore this complication and simply parameterize the lifetime
effects by a constant scattering rate $\Gamma$. We expect this approximation 
to be entirely adequate in the present context since the important qualitative
features of the tunneling conductance discussed below emerge clearly in the
clean limit $\Gamma=0$.

As alluded to in the Introduction, nontrivial dependence of the matrix element
$M_{\bf k}$ on the angle $\phi$ of the ${\bf k}$-vector 
on the Fermi surface has measurable consequences
for the tunneling conductance in an anisotropic superconductor. In tetragonal
cuprates band structure considerations imply strong anisotropy of the 
{\em interlayer} tunneling matrix element $t_\perp({\bf k})\propto \cos 2\phi$
\cite{wheatley1,andersen1}. One direct consequence of this anisotropy is
the well known qualitative difference between the temperature dependences of
the in-plane and $c$-axis penetration depths\cite{wheatley1}.
 It is reasonable to expect that
the structure of $t_\perp({\bf k})$ will directly translate into similar 
anisotropy in the matrix element $M_{\bf k}$ for tunneling between the 
superconductor and the STS tip.
The appendix confirms this expectation by providing a formal derivation of 
$M_{\bf k}$ from the transfer Hamiltonian formulation of the tunneling
problem. Motivated by these considerations 
in the following we study two models: a conventional model I with
$M_{\bf k}=M_0$ and an anisotropic  model II with $M_{\bf k}=M_2\cos 2\phi$,
as suggested by\cite{wheatley1}, where ${\bf k}=(k,\phi)$ in polar coordinates.
In model I tunneling conductance is simply proportional to LDOS while 
in model II tunneling from the zone diagonals is suppressed.

\begin{figure}[t]
\epsfxsize=9.5cm
\epsffile{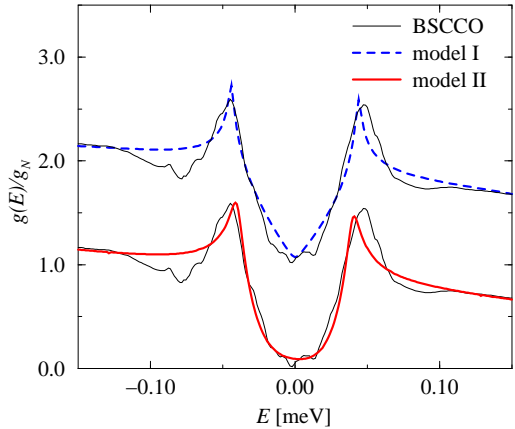}
\caption[]{Best fit of tunneling conductance from Eq.\ (\ref{cond1})
to the zero-field experimental data on underdoped BSCCO of Renner
et al.\cite{renner3}, for models I and II. The fitted parameters were
$\Delta_d=41.0$ meV and $\Gamma=1.4$meV for model I and  
$\Delta_d=39.7$ meV and $\Gamma=3.9$meV for model II.
A term linear in $E$ has been 
added to  Eq.\ (\ref{cond1}) in order to account for the background 
conductance and the data for model I are offset for clarity. 
}
\label{fig1}
\end{figure}

It is straightforward to numerically evaluate Eq.\ (\ref{cond1}) for the two
models and compare $g(V)$ to the experimental data. The result of the best fit 
to the zero-field data on BSCCO \cite{renner3} is displayed 
in Figure \ref{fig1}. It is seen that, as pointed out previously in Ref.\
\cite{millis1}, model II captures the qualitative features of the data 
much better than model I. The wide, U-shaped conductance near the zero bias 
appears to be a generic feature of the $c$-axis tunneling 
conductance in tetragonal cuprates\cite{renner3,wei1,ozy1,yazdani1}, and 
is inconsistent with 
linearly vanishing $g(V)$ of model I\cite{remark11}. We therefore conclude 
that the available
tunneling data are consistent with model II, as expected from the
band structure argument presented above.

\subsection{Finite field - semiclassical treatment}

The effects of applied magnetic field are taken into account by performing 
a semiclassical replacement 
\begin{equation}
\omega\to\omega-{\bf k}\cdot{\bf v}_s({\bf r})
\label{doppler}
\end{equation}
in the Greens function of Eq.\ (\ref{g1}). Here ${\bf v}_s({\bf r})$ is the 
local superfluid velocity
which in the vicinity of a single vortex has the form\cite{tinkham}
\begin{equation}
{\bf v}_s({\bf r})={\hat \theta\over 2mr},
\label{vs}
\end{equation}
and is cut off exponentially at distances in excess of the London penetration
depth $\lambda$. The semiclassical approximation (\ref{doppler}) was at the
heart of the original Volovik calculation of the specific heat\cite{volovik1}
and has been used
extensively to compute various spectral\cite{millis1,hirschfeld1,vekhter1} and 
transport\cite{hirschfeld2,vekhter2,franz2} properties of the mixed state. 
It appears to capture very well the essential physics of $d$-wave
quasiparticles moving on the slowly varying background of the vortex lattice. 

\begin{figure}[t]
\epsfxsize=9.5cm
\epsffile{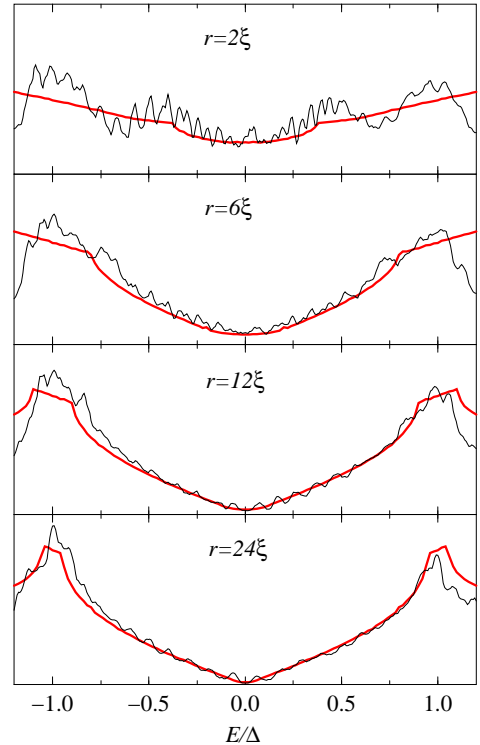}
\caption[]{Comparison of the LDOS obtained using semiclassical approximation
Eqs.\ (\ref{cond1}-\ref{vs}) with $\Gamma=0$ 
(thick lines) and the numerical solution 
of the BdG equations of Ref.\ \cite{franz3} (thin lines) at
indicated distances from the core. In both cases averages over the real-space
angle $\theta$ are plotted. 
}
\label{fig5}
\end{figure}

We have explicitly verified that, for $r\gtrsim 2\xi$,
 Eqs.\ (\ref{cond1}-\ref{vs}) yield very
reasonable LDOS profiles over the entire energy range of interest when 
compared to the results of a fully self-consistent calculation within
the BdG theory for a single $d$-wave vortex
\cite{franz3,remark1}. Figure \ref{fig5} illustrates the agreement between
the two approaches. Note in particular the excellent agreement in the
low energy part of the spectrum.
We believe that this comparison constitutes a stringent
test for the validity of the semiclassical approximation (\ref{doppler})
for the local quantities such as LDOS outside of the vortex core. 
As expected, however, inside the core the semiclassical approximation breaks
down, as visible in the top panel of Figure \ref{fig5}. In the strongly type-II
cuprates at fields well below $H_{c2}$ cores comprise only a small fraction of
the total volume. Semiclassical approximation thus works well almost 
everywhere which explains the success of the Volovik picture in modeling
of the mixed state. 

With the replacement (\ref{doppler}) the tunneling conductance (\ref{cond1})
becomes position dependent through the spatial dependence of the superfluid 
velocity. In the following we discuss the local 
tunneling conductance $g(V,{\bf r})$ near a single isolated vortex in a 
$d_{x^2-y^2}$ superconductor.

\section{Tunneling spectra in the vicinity of the vortex}

\subsection{Clean limit, zero temperature}

In the clean limit $\Gamma\to 0^+$ the Doppler-shifted 
spectral function assumes a simple form 
\begin{equation}
A({\bf k},\omega)=
\pi\sum_{\nu=\pm1}\left(1-\nu{\epsilon_{\bf k}\over E_{\bf k}}\right)
\delta(\omega-\eta+\nu E_{\bf k})
\label{spec1}
\end{equation}
where $E_{\bf k}=\sqrt{\epsilon_{\bf k}^2 + \Delta_{\bf k}^2}$ and 
$\eta={\bf k}\cdot{\bf v}_s({\bf r})$. We now evaluate the corresponding 
tunneling conductance given by Eq.\ (\ref{cond1}).
At $T=0$ the $f'$ factor 
becomes a $\delta$-function and the $\omega$-integral is trivial.
The remaining sum over ${\bf k}$ is replaced 
by an integral in the usual manner.  Assuming the free electron 
dispersion $\epsilon_{\bf k}=k^2/2m-\epsilon_F$ we may use the 
$\delta$-function
in (\ref{spec1}) to explicitly perform the integral over the energy variable
and obtain the tunneling conductance as a Fermi surface average of the form 
\begin{equation}
{g(V,{\bf r})\over g_N}=\int_0^{2\pi}{d\phi\over 2\pi} 
{\rm Re}\left[{\left({\sqrt{2}\cos2\phi}\right)^n 
|eV-\eta|\over\sqrt{(eV-\eta)^2-\Delta(\phi)^2}}\right].
\label{cond2}
\end{equation}
Here $g_N$ is the normal state conductance,  $n$ assumes values of 0 and 2
for models I and II respectively and we have restricted ${\bf k}$ in $\eta$ 
to the Fermi surface. We may thus write
\begin{equation}
\eta={\bf k}_F\cdot{\bf v}_s({\bf r})={\pi\over 2}
\Delta_d{\xi\over r}\sin(\theta-\phi),
\label{eta}
\end{equation}
with ${\bf r}=(r,\theta)$ and $\xi=v_F/\pi\Delta_d$ the coherence length.

For arbitrary bias $V$ the conductance (\ref{cond2}) must be
evaluated numerically. However, for $|eV|\ll\Delta_d$, the integral is 
dominated by the regions close to the four nodes of $\Delta(\phi)$ and 
can be evaluated, to an excellent approximation, by expanding to leading 
order in $\phi$ near the nodes. For the zero-bias conductance we thus
obtain 
\begin{equation}
{g(0,{\bf r})\over g_N}\simeq\sum_{l=1}^4\int_0^{z_l/2}{d\phi\over\pi}
\left({2\sqrt{2}\phi}\right)^n
{z_l\over\sqrt{z_l^2-(2\phi)^2}},
\label{cond3}
\end{equation}
where 
\begin{equation}
z_l={\pi\over 2}{\xi\over r}|\sin(\theta-\phi_l)|,
\label{zl}
\end{equation}
and $\phi_l=\pi(2l-1)/4$ are the nodal points. The integral implied by 
Eq.\ (\ref{cond3}) is elementary and yields
\begin{equation}
{g(0,{\bf r})\over g_N}\simeq
c_n\left({\xi\over r}\right)^{n+1} 
\left[ |\sin\tilde\theta|^{n+1} + |\cos\tilde\theta|^{n+1}\right],
\label{cond4}
\end{equation}
where $\tilde\theta=\theta-\pi/4$ is the polar angle measured from a node, 
$c_0=\pi/4\simeq 0.78$ and $c_2=\pi^3/16\simeq 1.94$. 

According to Eq.\ (\ref{cond4}) the symmetry of the tunneling matrix element
$M_{\bf k}$ has profound consequence for the spatial dependence of 
$g(0,{\bf r})$ near
the vortex core. Most importantly we notice that the decay with the distance 
$r$ from the core is much more rapid in model II, where 
$g(0,{\bf r})\sim (\xi/r)^3$, compared to the $\xi/r$ behavior in model I. For
instance, at $r=3\xi$ the zero-bias conductance will be suppressed by a factor
of 3 in model I but by a factor of 27 in model II. We argue that this 
difference is a very likely reason for the observed absence of $1/r$ tails
in STS measurements.  

\begin{figure}[t]
\epsfxsize=9.5cm
\epsffile{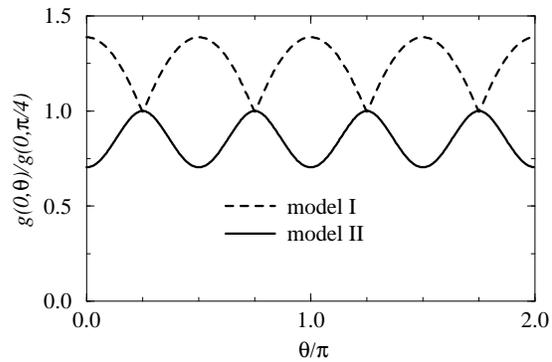}
\caption[]{Zero-bias tunneling conductance as a function of real space 
polar angle $\theta$ around the vortex normalized to unity along the nodal 
direction $\theta=\pi/4$.
}
\label{fig2}
\end{figure}
Eq.\ (\ref{cond4}) implies an interesting prediction for the 
the angular dependence of $g(0,{\bf r})$ for fixed $r$, corresponding to taking
a scan along a circle of the radius $r$ centered at the core. This  
angular dependence is a result of the underlying   
${\bf k}$-space anisotropy of the gap function and is 
illustrated in Figure \ref{fig2} for models I and II. 
While the ratio between maximum and minimum 
is $\sqrt{2}$ for the both models, we observe that there is a qualitative
difference between the two models in the positions of maxima and minima.
Based on our expectation that model II provides the correct description
of tunneling into BSCCO, we predict that $g(0,{\bf r})$ should exhibit minima
along the $\theta=0$ and maxima along $\pi/4$ and equivalent directions.
It would be most interesting to see if this expectation could be confirmed 
experimentally.

It is interesting to note that numerical calculations within the Eilenberger 
formalism\cite{maki1,ichioka2} show maxima of zero-bias LDOS along the
$\pi/4$ directions, in contradiction to the above conclusions. On the other 
hand, numerical
computations within the BdG formalism\cite{maki2}, 
when averaged over a narrow range
of energies near zero bias, show LDOS consistent with our results.

\subsection{Effect of temperature and finite lifetime}

It is evident that finite lifetime and temperature will affect the prediction
(\ref{cond4}) for the tunneling conductance, since they both lead 
to finite zero-bias conductance even in the absence of field. 
One expects that these effects
become significant at distances from the core  
where the characteristic Doppler-shift energy $E_D=\Delta_d(\xi/r)$ 
becomes comparable to $\Gamma$ or $T$. At 4.2K, which 
is typically the temperature of STS experiment, temperature broadening 
becomes important at $r/\xi\gtrsim\Delta_d/4.2$K which is a number 
of order 100 in cuprates. Thus, for low temperature tunneling the effects of
thermal broadening are unimportant, except at large distances from the cores.  
Scattering rate $\Gamma$, on the other hand, can be a significant
fraction of the maximum gap in cuprates, as evidenced by a relatively large
zero-bias conductance observed experimentally\cite{renner2,ozy1}, and will 
therefore cause significant broadening at distances of several $\xi$ from the
core. We now discuss the effect of finite lifetime $\Gamma$ in some detail.

For $\Gamma>0$ the spectral function becomes
\begin{equation}
A({\bf k},\omega)=
\sum_{\nu=\pm1}\left(1-\nu{\epsilon_{\bf k}\over E_{\bf k}}\right)
{\Gamma\over(\omega-\eta+\nu E_{\bf k})^2+\Gamma^2}.
\label{spec2}
\end{equation}
%
Evaluation of the ${\bf k}$-sum in Eq.\ (\ref{cond1}) is now somewhat more
involved since we no longer have a $\delta$-function at our disposal
to perform the energy integral. At $T=0$ the zero-bias 
tunneling conductance is
\begin{eqnarray}
g(0,{\bf r})&=&C\int_0^{2\pi}d\phi(\cos2\phi)^n\nonumber\\
&\times&\int_0^\infty d\epsilon
\left[{\Gamma\over(\eta+E)^2+\Gamma^2}+{\Gamma\over(\eta-E)^2+\Gamma^2}\right],
\label{condd1}
\end{eqnarray}
where $E=\sqrt{\epsilon^2+\Delta(\phi)^2}$ and $C$ contains all the constant 
prefactors. We evaluate (\ref{condd1}) by linearizing all the functions
in the integrand containing $\phi$ around the four nodes of the gap function, 
e.g. $\Delta(\phi-\phi_l)\approx 2\Delta_d\phi$,
and extending the angular integration to infinity. Under the assumption that 
$\Gamma, |\eta|\ll\Delta_d$ the resulting integral can be evaluated by 
making use of new variables, $u=\sqrt{\epsilon^2+(2\Delta_d\phi)^2}$ and 
$\alpha=\arctan[\epsilon/2\Delta_d\phi]$. The $\alpha$-integration is trivial
and we obtain
\begin{eqnarray}
{g(0,{\bf r})\over g_N}&=&{1\over 4\pi}\sum_{l=1}^4\int_0^\Lambda
du \left({u\over\Delta_d}\right)^{n+1}
\biggl[{\Gamma\over(u+\Delta_dz_l)^2+\Gamma^2}\nonumber\\
&& \ \ \ \ \ \ \ \ \ \ \ \ \ \ \ \ \ \ +
{\Gamma\over(u-\Delta_dz_l)^2+\Gamma^2}\biggr],
\label{condd2}
\end{eqnarray}
where $z_l$ is defined in (\ref{zl}) and
$\Lambda$ is a cutoff of the order of $\Delta_d$ imposed in order to assure 
convergence at large $u$\cite{remark2}. 
Finally, carrying out the integral and keeping in mind that
$\Gamma, |\eta|\ll\Delta_d$, we obtain the leading terms in model I,
\begin{equation}
{g(0,{\bf r})\over g_N}\simeq{1\over 4\pi}\sum_{l=1}^4
\left[2z_l\arctan\left({z_l\over\gamma}\right)-
\gamma\ln(\gamma^2+z_l^2)\right],
\label{conddI}
\end{equation}
and for model II,
\begin{equation}
{g(0,{\bf r})\over g_N}\simeq{1\over 4\pi}\sum_{l=1}^4
\biggl[ 2z_l(z_l^2-3\gamma^2)\arctan
\left({z_l\over\gamma}\right) + \gamma \biggr].
\label{conddII}
\end{equation}
Here we have set $\Lambda=\Delta_d$ and defined a dimensionless scattering
rate $\gamma=\Gamma/\Delta_d$. 

\begin{figure}[t]
\epsfxsize=9.5cm
\epsffile{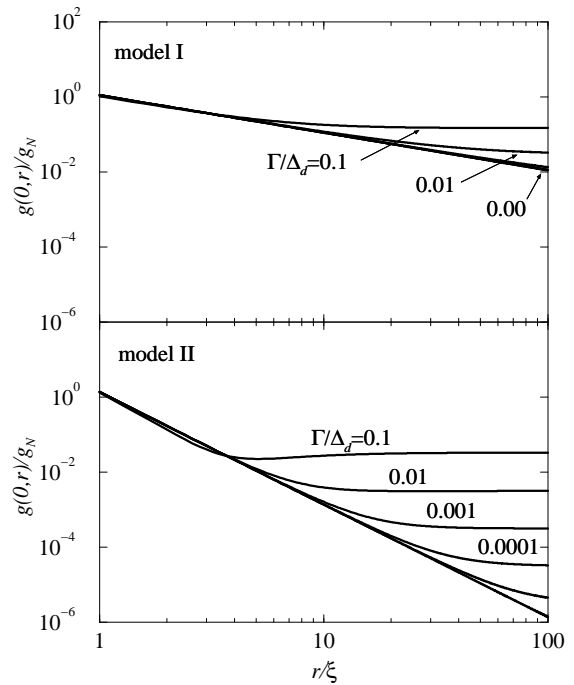}
\caption[]{Effect of finite quasiparticle lifetime $\Gamma$ on the amplitude 
of the zero-bias tunneling conductance as a function of distance $r$
from the vortex core (for $\phi=0$). Note the log-log scale. The straight 
lines with slopes -1 and -3 mark the expected asymptotic behaviors for 
$\Gamma=0$ in model I and II respectively. }
\label{fig3}
\end{figure}
Equations (\ref{conddI}) and (\ref{conddII}) exhibit the correct behavior
in the limit $\Gamma\to 0^+$ when compared to the result for the clean system
(\ref{cond4}). For finite $\Gamma$ they describe the crossover
from the $1/r^{n+1}$ behavior close to the core ($r\ll r_\Gamma$)
to the lifetime dominated constant zero-bias conductance far from the core
($r\gg r_\Gamma$). Inspection of
Eqs.\ (\ref{conddI}) and (\ref{conddII}) reveals that the dependence of the
crossover scale $r_\Gamma$ on $\Gamma$ is more subtle than one would expect 
from the simple argument
involving the Doppler-shift energy $E_D$ presented above. In particular
we find that 
\begin{equation}
r_\Gamma\sim -\xi/(\gamma\ln\gamma)
\label{rg1}
\end{equation}
for model I while
\begin{equation}
r_\Gamma\sim \xi/\gamma^{1/3} 
\label{rg2}
\end{equation}
for model II. In both cases lifetime effects
become important at {\em shorter} distances from the core than one would 
expect from the naive estimate $r_\Gamma\sim\xi/\gamma$. This is 
illustrated in Figure \ref{fig3} where we plot the crossover functions
for the tunneling conductance given by  Eqs.\ (\ref{conddI}) and 
(\ref{conddII}) as a function of distance $r$ from the core for a number
of lifetimes $\Gamma$. We note that in model II lifetime effects are much 
more efficient in destroying the clean power law behavior than in model I, 
consistent with Eqs.\ (\ref{rg1}) and (\ref{rg2}).

\begin{figure}[t]
\epsfxsize=9.5cm
\epsffile{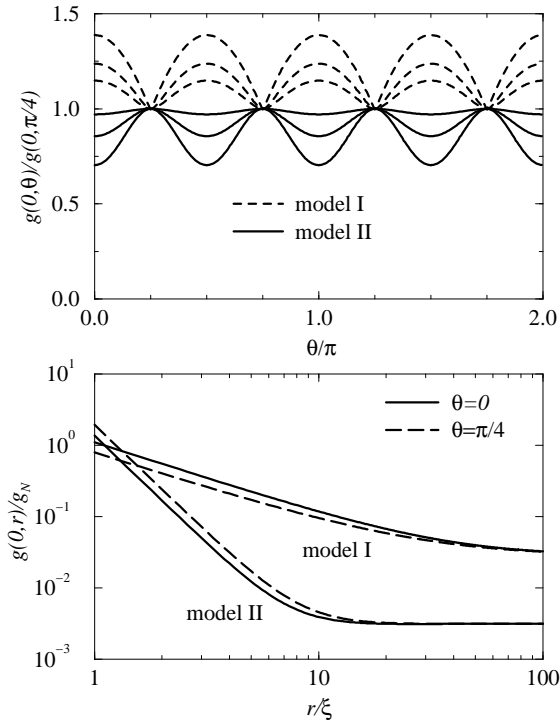}
\caption[]{Effect of finite quasiparticle lifetime $\Gamma$ on the angular
anisotropy of the zero-bias tunneling conductance. Upper panel: angular
dependence (normalized to unity at 
$\theta=\pi/4$) for $\Gamma/\Delta_d=0.01$ and
distances $r/\xi=1,\ 10,\ 20$ in the order of diminishing anisotropy.
Lower panel: maximum and minimum conductivities as a function of distance
$r$ from the vortex core for $\Gamma/\Delta_d=0.01$. }
\label{fig4}
\end{figure}

The lifetime effects will also affect the anisotropy of the tunneling
conductance around the core, since one would expect $g(0,{\bf r})$ 
to become isotropic
at distances larger than $r_\Gamma$. Figure \ref{fig4} displays the evolution
of the anisotropy as a function of angle $\theta$ 
for increasing distance $r$ at
constant $\Gamma$ (upper panel). Lower panel compares the angular maximum
to the minimum as a function of $r$. As expected, lifetime effects
wash out the anisotropy at distances from the core in excess of the 
crossover scale $r_\Gamma$. 

\section{Summary and conclusions}

Our main objective was to reconcile the apparent absence of the $1/r$ tails in 
the local tunneling conductance in the vicinity of a $d$-wave vortex,  
expected on the basis of a simple Volovik 
model, with the general consensus 
that such a semiclassical picture captures the essential 
physics of the mixed state in cuprates. We argued that the STS measurements 
are 
likely dominated by the non-trivial structure of the tunneling matrix element,
which derives from the well known band structure 
considerations\cite{wheatley1,andersen1}. 
We showed that, for the matrix element of the form $M_{\bf k}=M_2\cos(2\phi)$ 
(model II), in the absence of lifetime effects, the zero-bias tunneling 
conductance
power law is modified to $1/r^3$, making it vanish much faster than the $1/r$ 
tail obtained for constant $M_{\bf k}$ (model I). 

We predicted a substantial angular anisotropy of
the zero-bias tunneling conductance, 
with maximum to minimum ratio of $\sqrt{2}$, and with exactly opposite
arrangement of extrema in the two models. Our conclusion regarding 
the precise form of this anisotropy, however, is 
somewhat less certain in view of the fact that it disagrees with the 
numerical results obtained within the semiclassical Eilenberger formalism
\cite{maki1,ichioka2} (see however Ref.\ \cite{maki2}). Nonetheless, our
present results clearly indicate that the structure of the tunneling matrix
element will have significant effect on the real-space anisotropy of
$g(V,{\bf r})$. The reliable information about the details of this effect
can be obtained in a relatively straightforward manner 
by incorporating the non-trivial matrix element into a fully
self-consistent BdG calculation along the lines of Ref.\ \cite{franz3}; the
work on this is in progress. Within this method it will also be interesting
to study the effect of the matrix element on the tunneling spectra in the
vortex core, a problem inaccessible to the semiclassical approximation.  

Lifetime effects will cut off both the power law decay and
the anisotropy at distances beyond the crossover length scale 
$r_\Gamma$ given by Eqs.\  (\ref{rg1}) and (\ref{rg2}), where $r_\Gamma$ is 
always shorter in model II for a given value of $\Gamma$. This means that if  
model II is the physically relevant one, as it appears to be the case for 
BSCCO, then the lifetime effects may likely render the experimental detection
of the asymptotic $1/r^3$ behavior very difficult. For instance, if 
we assume $\Gamma/\Delta_d=0.1$, which is physically reasonable for 
BSCCO at low $T$\cite{millis1}, then Eq.\ (\ref{rg2}) implies 
$r_\Gamma\simeq 2.1\xi$, meaning that there will be virtually no asymptotic
region where the $1/r^3$ behavior could be observed. 
Instead, one would see an onset of the 
constant zero-bias tunneling conductance just
outside of the vortex core. This theoretical conclusion is in fact consistent
with the STS data of Renner et al.\cite{renner2}. Even in the presence 
of substantial lifetime effects it should still be in 
principle possible to observe the predicted angular anisotropy of 
$g(0,{\bf r})$
close to the the core. The existing experimental data on BSCCO in fact show
a definite hint of a fourfold anisotropy, which is, however, discernible only
at high bias\cite{fischer1}.
The above analysis suggests that the asymptotic tails
might become observable in the cleanest samples characterized by low
value of the zero-bias conductance in the absence of field. Clearly,
the effect would be difficult to discern in the existing data on 
YBCO\cite{renner1} which display large zero-bias conductance of unknown
origin.  

It would thus appear that the 
proper inclusion of the anisotropy in the tunneling
matrix element naturally resolves the conflict between the STS and 
heat capacity measurements, since the latter 
is obviously insensitive to the structure of the matrix element. While certain
details still await experimental verification, this seems to be a 
satisfactory tentative 
conclusion, especially since additional experimental evidence emerged 
recently that appears to further solidify the support for the Volovik-type
description of the mixed state in cuprates. In particular
measurements of the complex conductivity (using coherent terahertz 
spectroscopy) in the mixed state of BSCCO films reported by 
Malozzi et al.\cite{malozzi1} 
were claimed to be consistent with the picture of the
Doppler-shifted local quasiparticle spectra\cite{remark12}. 
Although somewhat
less directly the experiments on thermal transport in the mixed state of
cuprates\cite{krishana1,may1} also seem to support the basic picture of 
a Doppler-shifted local quasiparticle in that they are consistent with 
theoretical
models based on this picture\cite{hirschfeld1,franz2}. We may thus conclude 
that by invoking the properties of the tunneling matrix element and lifetime 
effects the existing tunneling data indeed can be brought to agreement with 
other experiments and that 
the overall picture of the mixed state in cuprates appears consistent with 
that of a well defined quasiparticle propagating on a  background of 
slowly varying supercurrents in the vortex array.

\acknowledgments	
The authors are indebted to \O. Fischer, Ch. Renner and A. J. Millis for 
stimulating discussions.  This research was supported by NSF
grant DMR-9415549.

\appendix
\section*{c-axis tunneling conductance in cuprates}

According to the standard many-body formulation\cite{mahan1} the tunneling
current at a bias $V$ between a superconductor and a normal metal is given by
\begin{eqnarray}
I(V)&=&2e\sum_{{\bf k},{\bf p}}|T_{{\bf k}{\bf p}}|^2\int_{-\infty}^\infty
{d\epsilon\over 2\pi}A_S({\bf k},\epsilon)A_N({\bf p},\epsilon-eV)\nonumber\\
&& \ \ \ \ \ \ \ \ \ \ \ \ \  \ \ \ \times[f(\epsilon)-f(\epsilon-eV)].
\label{ai1}
\end{eqnarray}
Here $A_S$ and $A_N$ are spectral functions of the superconductor and the
normal metal respectively, $T_{{\bf k}{\bf p}}$ is the tunneling matrix element
and $e$ is electron charge. In the following we make the usual assumption
that the normal metal in the STS tip can be described 
by the spectral function
\begin{equation}
A_N({\bf p},\epsilon)=2\pi\delta(\epsilon-\xi_{\bf p}),
\label{ans}
\end{equation}
with a simple free electron 
dispersion $\xi_{\bf p}=({\bf p}_\perp^2+p_z^2)/2m-\epsilon_F$ where we have 
resolved ${\bf p}$ into components parallel ($p_z$) and perpendicular
(${\bf p}_\perp$) to the $z$ axis along which the tunneling current flows.  
The $\delta$-function in $A_N$ can be used to perform the $\epsilon$-integral
in (\ref{ai1}) to obtain
\begin{equation}
I(V)=2e\sum_{{\bf k},{\bf p}}|T_{{\bf k}{\bf p}}|^2
A_S({\bf k},\xi_{\bf p}+eV)[f(\xi_{\bf p}+eV)-f(\xi_{\bf p})].
\label{ai2}
\end{equation}

The expression for the tunneling matrix element can be derived from the 
transfer Hamiltonian formalism\cite{wolf1}:
\begin{equation}
|T_{{\bf k}{\bf p}}|^2=\left|{\partial\epsilon_{\bf k}\over\partial k_z}\right|
\left|{\partial\xi_{\bf p}\over\partial p_z}\right|
D(\epsilon_z)\delta({\bf k}_\perp-{\bf p}_\perp).
\label{at}
\end{equation}
Here $\epsilon_{\bf k}$ denotes the normal-state dispersion in the 
superconductor,
$D(\epsilon_z)$ is the barrier transmission coefficient ($\epsilon_z$ is
the energy of the electron inside  the barrier), and the $\delta$-function
reflects conservation of the perpendicular momentum under the conditions
of specular transmission. The latter can be used to carry out the 
summation over the transverse momentum ${\bf p}_\perp$ in Eq.\ (\ref{ai2}). 
The remaining sum over $p_z$ can be  converted into an integral in the usual 
manner.  The integral is then  transformed by making a 
substitution $\omega=(p_z^2+{\bf k}_\perp^2)/2m-\epsilon_F+eV$ and noting that
 the term $|\partial\xi_{\bf p}/\partial p_z|$ in $|T_{{\bf k}{\bf p}}|^2$ 
is precisely the Jacobian of this transformation; we obtain
\begin{equation}
I=2e\sum_{\bf k}\left|{\partial\epsilon_{\bf k}\over\partial k_z}
\right|D(\epsilon_z)
\int_{-\infty}^\infty{d\omega\over 2\pi}
A_S({\bf k},\omega)[f(\omega)-f(\omega-eV)].
\label{ai3}
\end{equation}
Differentiating with respect to $V$ we finally arrive at the expression for 
the funneling conductance
\begin{equation}
g(V)=-\int_{-\infty}^\infty{d\omega\over 2\pi}f'(\omega-eV)
\sum_{\bf k}|\tilde M_{\bf k}|^2A_S({\bf k},\omega),
\label{ag1}
\end{equation}
with $|\tilde M_{\bf k}|^2=e^2|\partial\epsilon_{\bf k}/\partial k_z|
D(\epsilon_z)$.

Although this last expression bears similarity to Eq.\ (\ref{cond1}),
the last term still contains summation over the 3D wave-vector ${\bf k}$ and it
therefore reflects the full 3D band structure of the superconductor. 
In the following we shall specialize to quasi-2D cuprates and 
simplify (\ref{ag1}) further by eliminating the $k_z$ summation, thereby
expressing the tunneling conductance in terms of $ab$-plane properties only. 
To this end we consider the single particle dispersion $\epsilon_{\bf k}$ of
the form deduced for tetragonal cuprates by Xiang and Wheatley\cite{wheatley1}:
\begin{equation}
\epsilon_{\bf k}=\epsilon_{\bf k}^\perp-t_z({\bf k}_\perp)\cos k_z,
\label{adis}
\end{equation}
where $\epsilon_{\bf k}^\perp$ is the $ab$-plane dispersion depending only
on ${\bf k}_\perp$ and 
\begin{equation}
t_z({\bf k}_\perp)={t_z\over4}(\cos k_x-\cos k_y)^2
\label{atz}
\end{equation}
is the inter-plane tunneling matrix element. In the following
we model
$\epsilon_{\bf k}^\perp$ by a free electron dispersion ${\bf k}_\perp^2/2m-
\epsilon_F$
and noting that only values of $t_z({\bf k}_\perp)$ close to the the Fermi 
surface are important we approximate
$t_z({\bf k}_\perp)\approx \tilde t_z\cos^22\phi$. While neither of these
two approximations is essential for the final result, adopting them will
greatly simplify the calculations. We may now convert the ${\bf k}$-sum in 
Eq.\ (\ref{ag1}) into an integral of the form (suppressing various
constant prefactors),
\begin{equation}
\int_{-\infty}^\infty dk_z\int_{0}^{2\pi}d\phi\int_{0}^\infty k_\perp d k_\perp
|\tilde M_{\bf k}|^2A_S(\epsilon_{\bf k},\Delta;\omega),
\label{aint1}
\end{equation}
where we have explicitly acknowledged the fact that $A_S({\bf k},\omega)$ 
will only
depend on the momentum variable through $\epsilon_{\bf k}$ and $\Delta(\phi)$.
We now convert the $k_\perp$ integral into an integral over the energy,
noting that, according to our assumptions, only the $\epsilon_{\bf k}^\perp$ 
component of $\epsilon_{\bf k}$ depends
on $k_\perp$ and thus the Jacobian of this transformation is simply a 
constant.  We thus arrive at the final result for the tunneling conductance
\begin{eqnarray}
g(V)&=&-\int_{-\infty}^\infty{d\omega\over 2\pi}f'(\omega-eV)
\nonumber\\
&&\ \ \ \  \times \int_{0}^{2\pi}d\phi\int_{-\infty}^\infty 
d\epsilon
|M(\phi)|^2A_S(\epsilon,\Delta;\omega),
\label{ag2}
\end{eqnarray}
where
\begin{equation}
|M(\phi)|^2={2e^2m\over (2\pi)^2}\int_{-\infty}^\infty {dk_z\over 2\pi}
\left|{\partial\epsilon_{\bf k}\over\partial k_z}\right|D(\epsilon_z)
=M_2^2\cos^22\phi,
\label{am}
\end{equation}
and $M_2$ is a constant.
By converting the $\epsilon$ and $\phi$ integrals in Eq.\ (\ref{ag2}) back to 
a 2D ${\bf k}$-vector sum we recover the expression for the tunneling 
conductance written down on purely intuitive grounds in Eq.\ (\ref{cond1}).

Superficially the result (\ref{ag2}) appears to contradict the 
conventional wisdom that the band structure effects
are ``invisible'' to tunneling because of the exact cancellation between the 
$|\partial\epsilon_{\bf k}/\partial k_z|$ factors in the $T_{{\bf k}{\bf p}}$ 
matrix and
the like factors resulting from the variable change $k_z\to\epsilon$ in the
integral\cite{wolf1,cancel}. In particular one could argue
that had we done the $k_z$-integral in Eq.\ (\ref{aint1}) first, such a
cancellation would have indeed occurred, leading to the result with 
$\phi$-independent matrix element.
However, in the present case we note that the appropriate 
$k_z\to\epsilon$ variable change would be illegitimate because of the 
divergence in the corresponding Jacobian for ${\bf k}_\perp$ coinciding with 
one
of the four nodes of $t_z({\bf k}_\perp)$. Therefore, as pointed out previously
in connection with tunneling in cuprates\cite{wei1}, the tunneling conductance
in fact {\em is} sensitive to non-trivial features in the band structure,
such as the matrix element that vanishes along certain ${\bf k}_\perp$ 
directions.


\begin{references}
\bibitem{remark1} Detailed comparison between the numerical work within
BdG formalism and the semiclassical approximation will be
given elsewhere.
\bibitem{remark2} The $u$-integral formally diverges at large values. This is
an artifact introduced by extending the angular integration to infinity. 
Numerical evaluation of Eq.\ (\ref{condd1}) confirms that imposing a cutoff
of the order of maximum gap $\Delta_d$ results in a very accurate approximation
as long as both $\Gamma$ and $|\eta|$ are small compared to  $\Delta_d$.
\bibitem{remark12} A puzzling
aspect of this data is that the effect persists up to 50K, where it should be
washed out by thermally generated quasiparticles and by increased inelastic
scattering rate which becomes comparable to the gap itself.
\end{references}
\end{document}